\documentclass[a4paper]{spie}  

\usepackage[]{graphicx}

\title{Development of super broadband interferometer in FIR} 

\author{Izumi S. Ohta\supit{a}, Makoto Hattori\supit{b} and 
Hiroshi Matsuo\supit{a}
\skiplinehalf
\supit{a}National Astronomical Observatory, Japan\\
\supit{b}Astronomical Institute, Tohoku University, Japan
}

\authorinfo{I. S. Ohta: E-mail: izumi.ohta@nao.ac.jp, Telephone: +81-422-34-3875, Fax:  +81-422-34-3864, Address: National Astronomical Observatory 2-21-1 Ohsawa Mitaka Tokyo 181-8588, Japan\\}

\begin{document} 
  \maketitle 

\begin{abstract}
We are developing the super broad band interferometer by applying the
Fourier Transform Spectrometer(FTS) to aperture synthesis system in mm and
sub-mm bands. We have constructed a compact system based on the Martin
and Puplett type Fourier Transform spectrometer (MP-FT). 
We call this equipment Multi-Fourier Transform interferometer
 (MuFT).
The band width of the system can be extended as large as one wants
contrary to the severely limited band width of the usual interferometer 
due to the speed of the AD converter. 
The direct detectors, e.g. bolometer, SIS video detector, 
can be used as the focal plane detectors. This type 
of detectors have a great advantage in FIR band since they are free
from the quantum limit of the noise which limits the sensitivity of
the heterodyne detectors used in the usual interferometers.  
Further, the direct detectors are able to make a large format array
contrary to the heterodyne detectors for which construction of a large format
array is practically difficult. These three characteristics make one be
possible to develop high sensitive super broad band FIR
interferometer with wide field of view. In the laboratory experiments,
we have succeeded in measuring the spectroscopically resolved
2D image of the source in 150GHz-900GHz band. The future
application of this technique to the observations from the space 
could open new interesting possibilities in FIR astronomy. 
\end{abstract}



\section{INTRODUCTION}

Sub-mm wave band is one of the yet unexplored wavebands 
in Astronomy although a lot of preliminary researches
have been attempted. One part of the difficulties for performing
observations in this band is owing to a large 
atmospheric absorption optical depth. 
Since sub-mm wave band is a boundary of the radio and
infrared astronomy, there remain a lot of frontiers in the 
development of the fundamental observational technology. 

If an interferometer which is able to use direct detectors as focal 
plane detector were developed, 
it could open a new horizon for mm and sub-mm 
wave astronomy since it shares both merits of single dish and interferometer. 
In this research, we have developed one of the concrete examples 
of such kind of new instruments by applying 
the aperture synthesis technique to the Fourier Transform Spectrometer (FTS). 
The application of the FTS to the aperture synthesis was first 
independently proposed by Itoh \& Ohtsuka(1986)\cite{Itoh} and Mariotti \&
Ridway (1988)\cite{Mariotti} in NIR. The system was referred to as double Fourier
interferometer by these authors since the obtained signal by the system
is Fourier transformation of both spectra and intensity distribution
on the sky. In these works, the Michelson type FTS was applied to the 
aperture synthesis.  

We considered the application
of the Martin \& Puplett type Fourier Transform spectrometer\cite{Ma-Pu}
(hereafter MP-FT) to the aperture synthesis system in wide band\cite{sen}.
In this system, wire grid is used as beam splitter.
By setting the wire grids appropriately, 2D intensity distribution 
of four Stokes parameters are able to be measured by our instrument.
The obtained signal by our system 
is Fourier transformation of spectra and 
intensity distributions of  
4 Stokes parameters (multiple components) on the sky. 
Therefore, we refer this system as Multi-Fourier Transform interferometer 
abbreviated 
to MuFT. 
The abbreviated name of MuFT also contains the meaning that  
this instrument measures Mutual correlation of the source signal instead of 
auto-correlation as in the case of usual FTS
although it is based on FTS.  
This is the first work which studied the application of 
the aperture synthesis technique to the MP-FT.  
There was also no works which concretely studied the application 
of the aperture synthesis technique to the FTS including the Michelson type 
in mm and sub-mm wave band. 

Applying this technique to space born mission is one of the best 
possibilities to extract the maximum ability of MuFT since 
there is no restriction on the band width from the atmospheric absorptions. 
Mather et al. \cite{Mather} has been proposing space born 
Far Infrared Observatory based on this kind of technique to NASA. 

\section{Fundamentals of Multi-Fourier Transform interferometer}
\label{sec:system}

The MuFT is a system which makes possible the imaging and
spectroscopy in a wide band by combining Wiener-Khinchine Formula and
 van Citter Zernike Formula\cite{sen}. By setting the wire grids appropriately,
acquisition of polarization information is possible. 
In this section, fundamentals of the components of the MuFT system and
fundamentals of the observables by the MuFT are summarized. 
The advantages of the MuFT are also summarized. 

\subsection{Components of MuFT}
   \begin{figure}
   \begin{center}
   \begin{tabular}{c}
   \includegraphics[height=5cm]{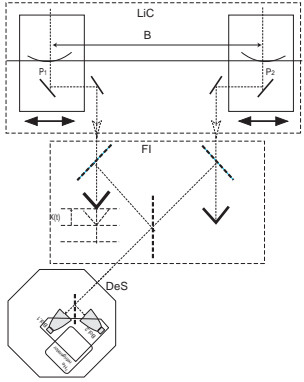}
   \end{tabular}
   \end{center}
   \caption[MuFT]{\label{fig:MuFT0}Simplified schematic diagram of MuFT. }
\end{figure} 

Fig.\ref{fig:MuFT0} is a schematic diagram of MuFT. 
From top of the page, there are Light Concentrating part(LiC), Fourier 
Interference part(FI) and  Detection \& Sampling part(DeS). 
The detected signal is transfered to computer after passing 
the electrical amplifier, noise filter and AD converter. 
The transferred data are analyzed by the appropriate data 
analysis system. 
Since the data is transformed into information of Polarimetry, 
Spectroscopy and Image in the data analysis system, 
the data analysis system is referred to PSI. In the followings, details of each components and present resources are explained.

LiC part first performs a division of wave front 
of the incident source signal,  and guides the signal to the Fourier 
interference part. The interval and orientation of the baseline 
vectors, and pointing the objects are also controlled in this part. 
The antenna in usual radio interferometer and siderostat in optical
interferometer  are corresponding to this part. 

In the case of using grounded telescopes,
a large geometrical light path difference must be corrected by using 
long delay line.  In our current system, the geometrical light path difference 
is canceled out by controlling the position of each telescopes so that 
the reference point of the targeted source is kept to be pointed along a 
bisector of the baseline vector perpendicular to the baseline vector. 
The same reference point must be on the center of the FOV.
 
As one of the example, we are currently using a heliostat as pointing 
and guiding system.
It guides the light from the targeted source so that 
the incident light always comes down from the vertical direction 
for the observer who is sitting below the second mirror. 
Put the optical elements for performing division of wave front
on the table below the second mirror, and perform the division of 
wave front. 
The baseline interval is controllable by controlling 
the positions of these optical elements. 
When very high spatial resolution is desired,   
using heliostat is not realistic solution since the base line interval is
limited by the size of the heliostat. 
Something like two mirrors connected by one rail mounted on turn table
could be a possible solution like as SPIRIT\cite{Mather1}. 
In the space, using multi-satellites could be an attractive solution. 

   \begin{figure}
   \begin{center}
   \begin{tabular}{c}
   \includegraphics[height=5cm]{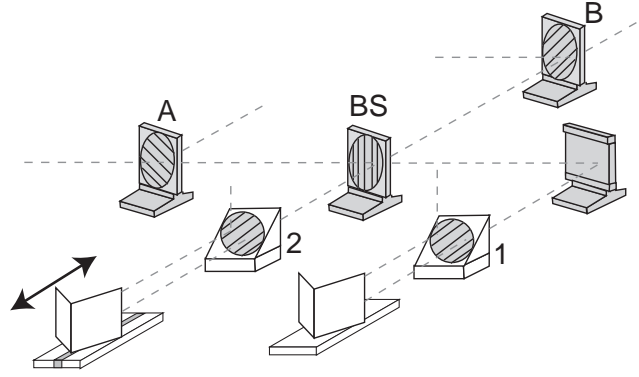}
   \end{tabular}
   \end{center}
   \caption{\label{fig:FI}Fourier Interference part of MuFT. This illustrates one of the possible configurations of WGs which observes Stokes parameters of I and Q. The example of arrangement of WG of Option.1. A grid-like elements are WGs. Dashed lines represent paths of light beams.  Light beams are guided by LiC part to the WG 1 and 2 from top to down vertically. Polarization components parallel to wires of the WG 1 and 2 are reflected toward the roof top mirrors (RTM). Other components are transmitted toward the down stairs. The reflected components are visiting WG 1 and 2 again and are transmitted through WG 1 and 2 toward BS WG this time. Light path length of one of the reflected components are modulated by moving the RTM. At WG BS, two combined beams are generated as illustrated in the figure. The combined beams are composed by two orthogonally polarized beams. Each of the orthogonal components corresponds to the light beams reflected by WG 1 and 2. To take a project!
 ion components towards the same polarization direction of these orthogonally polarized beams, WG A and B are put in front of the detectors. Otherwise, the mutual interference of the light beams taken by WG 1 and 2 does not observe.}
\end{figure}

The Fourier Interference part is the instrument in which the divided 
light beams are recombined after modulating the one of the light path 
length and the hart of the MuFT system. 
It has two entrance windows to get two light beams guided 
from the light concentration part. 

Wire grid polarizers are used as beam splitters and beam combiners. 
One of the example of the FI part is illustrated in Fig.\ref{fig:FI}. 
Depending on the alignment of the wire grids 1 and 2, observables of the FI part is different. There are two basic alignment of the wire grids.
When the direction of the wires of WG 1 and WG 2 are parallel as illustrated in Fig.\ref{fig:FI}, Stokes parameters of I and Q are observables. This is referred to option 1. When the direction of the wires of WG1 and WG2 are orthogonal, Stokes parameters of U and V are observables. This is referred to option 2.  
By treating incident waves as vector waves, 
interferogram of each option can be expressed as follows, 
\begin{eqnarray}
\label{eq:Op12-2}
I_{Op1}&\propto &\int_{\Omega}\int
d\nu\frac{\mathcal{I}(\vec{\theta},\nu)\pm \mathcal{Q}(\vec{\theta},\nu)}{2}
\left[1+\cos2\pi \frac{\nu}{c}(\vec{b}\cdot \vec{\theta }+2x)\right],
\nonumber \\
I_{Op2}&\propto &\int ds \int d\nu {1\over 2}\left[\mathcal{I}(\vec{\theta},\nu
  )\right. \nonumber \\
&&\left. + \left(\mathcal{U}(\vec{\theta},\nu )\cos[2\pi \frac{\nu}{c}
    (\vec{b}\cdot \vec{\theta } +2x)]
\pm \mathcal{V}(\vec{\theta},\nu)\sin[2\pi \frac{\nu}{c}(\vec{b}\cdot \vec{\theta
  } +2x)]\right) \right], 
\end{eqnarray}
where $\pm$ corresponds how the direction of the WG1 and 2 are set. 
These results show that the MuFT is possible to measure all
intensity distributions of 4 Stokes
parameters  for each frequency in wide band, eg.  
$\mathcal{I}(\vec{\theta},\nu)$, $\mathcal{Q}(\vec{\theta},\nu)$,
$\mathcal{U}(\vec{\theta},\nu)$, $\mathcal{V}(\vec{\theta},\nu)$.
Since only three of 4 Stokes parameters are independent for 
completely polarized waves, 
systematic measurements errors could be self calibrated 
in MuFT by measuring all of 4 Stokes parameters for the completely polarized light beams. 
For the intensity measurement, the observations by the option 1 is enough. 

   \begin{figure}
   \begin{center}
   \begin{tabular}{c}
   \includegraphics[height=5cm]{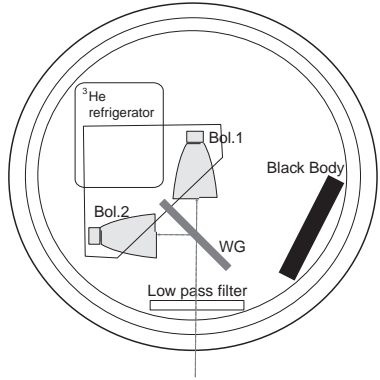}
   \end{tabular}
   \end{center}
   \caption{\label{fig:DetS}Detection system of MuFT. }
\end{figure}

The DetS part which is optimized to MuFT, is illustrated in Fig.\ref{fig:DetS}. The most of the works on this 
developments were performed by Hamaji and summarized in his master 
thesis at Tohoku University.  The difference of reflecting and transmitting 
signals by the output wire grid reproduces 
full mutual coherence function and removes the unwanted DC 
components. The DC fluctuations originated from the atmospheric fluctuation
and the thermal fluctuation of the detector systems are common both for 
the transmitted and reflected components. Therefore, these fluctuations are 
expected to be removed out by taking the difference of both signals.  

The PSI system performs control of MuFT, and analysis of acquainted data.
Control of equipment is mainly built based on LabView. Specifically, 
control of source tracking and monitor of a baseline length vector, 
an operation mirror, control of direction of  WG, 
acquisition of data are performed. 
For the analysis of acquisited data, it is built mainly based on IDL. 
Fourier transformation of the acquisited data in each 
baseline length vector is performed and processing using adjustment 
and the reference of data is performed and the data based on the 
whole data is compounded.

\subsection{Advantages of MuFT}

\paragraph{High sensitivity and wide band width interferometer in FIR}

This system can use direct detectors as the focal plane detectors of the 
interferometer, like bolometer or Super-conductive Tunnel Junction 
detectors(STJ).  They are the most  sensitive detectors in
mm, sub-mm and FIR wave bands since they are free from quantum limit 
of the noise which limits the sensitivity of the heterodyne detectors 
used in the usual interferometers in these wave bands.
They also make possible very wide band observations. 
The band width of the system can be extended as large as 
one wants contrary to the severely limited band width of 
the usual interferometers which use heterodyne detectors. 
Since the wire grid has an uniform reflection and transmission rates
for all frequencies in mm, sub-mm and FIR,  
the MuFT is suited to broad band observations. 

\paragraph{Wide field of view}

Constructing large format heterodyne spatial arrays are not yet feasible, 
since generator of the reference signal and associated electric circuit 
required to be mounted for each detectors
are bulky and not amenable to automated production that uses
integrated circuit technology. 
Therefore, yet a single detector is used in radio interferometers,
including ALMA.  It severely limits a FOV of the instruments.
On the other hand, a large form spatial arrays of direct detectors 
are already operating.  Developments of new type of direct detectors 
aiming a dramatically extension of  the pixel number have 
been also progressed.   
Since MuFT can use direct detectors as the focal plane detector
of interferometer, it dramatically extends the FOV of the mm and
sub-mm interferometers. 
To develop this technique, a group in NASA is now performing
laboratory experiments by using optical CCD. 
They refer this experiment Wide-field Imaging Interferometry Testbed
(WIIT)\cite{Leisawitz}. 

\paragraph{Large dynamic range}

When source intensity distribution is independent from the frequency, 
we can sum up all images obtained for different frequencies 
into single image. In other ward, a single baseline observation 
is equivalent to the various baseline interval observations. 
Long wave length see the object with low resolution. Short wave 
length see it with high resolution. Therefore, large dynamic 
range is retained efficiently.

\paragraph{Linear and Circular Poralimetry}
The MuFT can measure all 4 components of Stokes parameters 
\cite{hattori}. 

\paragraph{etc.}
Since this is an interferometer, it is only sensitive to interference
signal, in other word coherence signal.  
Spatial scale of the atmospheric fluctuation usually has a long 
wave length. Therefore, interferometer is usually insensitive for 
atmospheric fluctuation, in other word it is strong against 
the atmospheric fluctuation.

\section{Experiments in Laboratory} 

   \begin{figure}
   \begin{center}
   \begin{tabular}{c}
   \includegraphics[height=5cm]{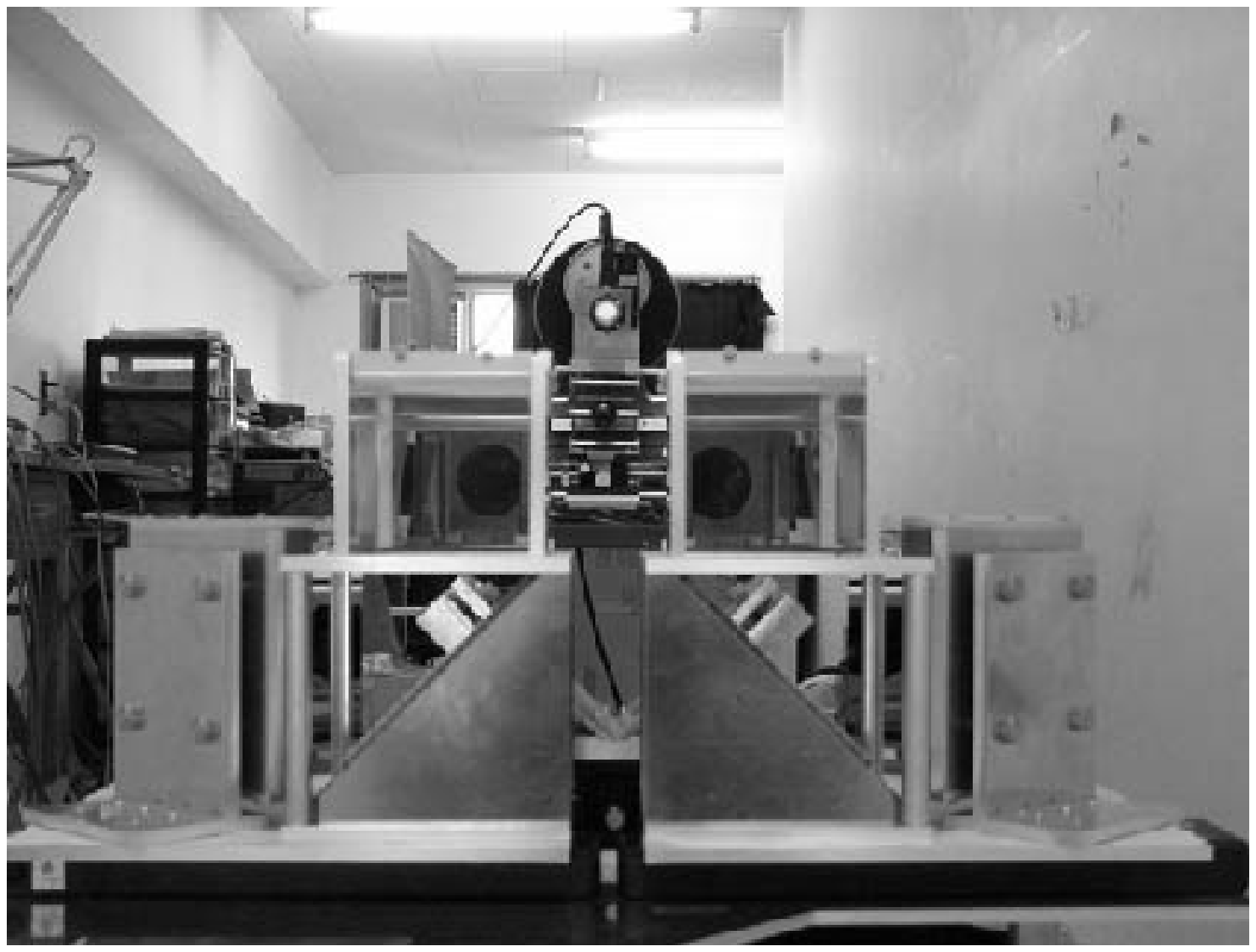}
\includegraphics[height=5cm]{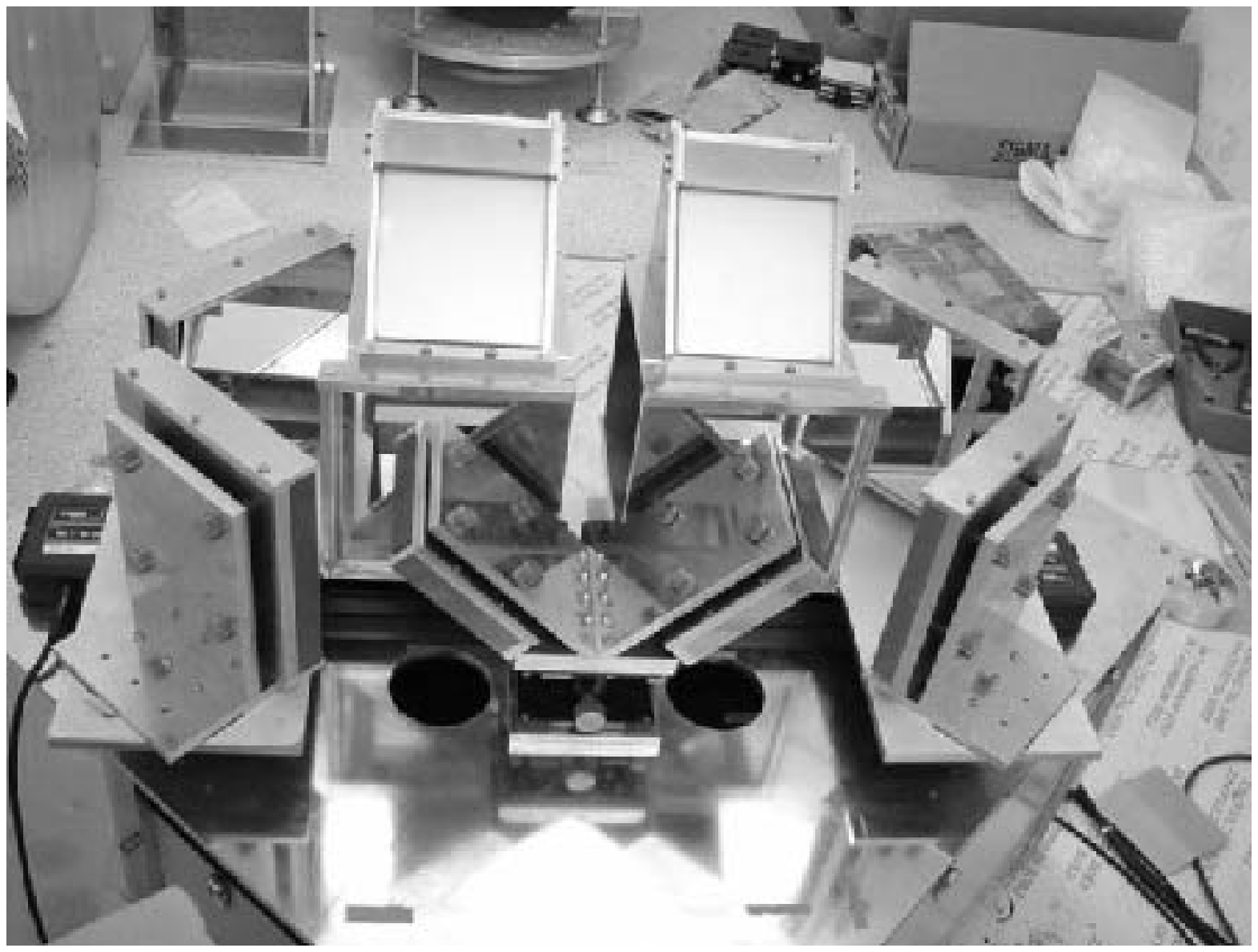}
   \end{tabular}
   \end{center}
   \caption{\label{fig:optics}Optical system of Labo. experiment. }
\end{figure}

The results of the laboratory experiments conducted with the prototype 
FI part of MuFT, FT-0.1. We have been performing experiments in the 
laboratory aimed to verify the theoretical considerations of the MuFT 
and quantify the abilities of the MuFT. The two dimensional extended source imaging was performed using the black body 
emitter as the source. The parameters of these experiments are summarized in 
Table.\ref{2dP}.

Light source is 1000 centigrade black body. 
As the collimator, a spherical mirror with a diameter of 80cm 
and curvature radius $R=640$cm was used. 
The mirror is made of aluminum and coated with sapphire about 0.2$\mu$m in 
depth. The transmission of the Sapphire coating 
is high against a submillimeter wave. 
The thickness of coating is small enough compared with wavelengths 
which we were interested in. Therefore it is not necessary to take the 
effect of absorption into consideration. Four trapezoid masks pattern are 
put in front of the black body aperture which are referred to A-D. 
The mask is made of aluminum and is handmade. 
An optical system is shown in  Fig.\ref{fig:optics}. 
Baseline length is varied from 18 to 28 cm for mask A and from 
18 to 38 cm for masks  B - D. These experiments were conducted in August, 
2003(A), and February, 2004(B - D). 

Firstly, Mercury Lamp  
was used for optical-axis adjustment since the detection of the mutual coherence signals 
of mercury lamp is much easier than that of black body.
The optical-axis adjustment is performed by following procedure.

(1) Optical-axis is adjusted by using the laser beams of
 the sub interferometer in FI-0.1 at first. The source was put on
the focus of the laser beams  
which reverses light path.

(2)Then, check by eyes whether the source is put on the appropriate position.    

(3)And, height and distance of the source 
were adjusted so that amplitude of the source signals guided 
through each entrance WG (WG 1 and 2)  
by  closing one of two light paths.  

(4) The source position is searched at where 
the mutual coherence signal becomes maximum. 
Check whether the peak position of mutual coherence signals 
in the interferogram does not change 
as increasing the baseline length. 

If a peak position moves  as changing baseline length,   
the direction of the incident beam is inclined 
from the direction normal to the baseline
and the source is not put on the center of the FOV. 
The procedures from (1) to (4) were performed iteratively several times. 
Since it was the optical system of off-axis of the collimator mirror, 
the black body source was inclined so that the plain of the exit window 
of the black body source is normal to the optical axis of the collimator
 mirror.
The experiment was performed with an assumption that rotation of mask 
is equivalent to rotation in baseline plane.  The rotation of the 
black body radiator by synchronizing the rotation of the mask is hard. 
So the black body is fixed. Since the surface intensity distribution 
of the black body radiator can be regarded as uniform and the emission 
from the different part of the black body can be regarded as incoherent, rotation of the mask 
can be treated as equivalent to the rotation of the source and the rotation 
in the baseline plane. In order to carry out imaging, 
it is necessary to take data at various $u, v$ points. 

Baseline length sampled 18 to 28cm with 2cm spacing (set A)
or 18cm to 38cm with 12cm spacing (set B).
The mask rotation angles were changed with 15 or 30 degrees intervals.
The number of the $u-v$ sampling points are 54 points in the case of set A 
and 18 points in the case of set B.
The measurements were carried out with an operation mirror speed of 
0.8 cm/s, and 20 or 10 scans per one sampling point. 

The angle of rotation of baseline length and a mask was controlled by the 
number of pulses sent to an automatic x and rotation stage.
The optimal operating temperature of the bolometer used for the 
experiment 
was 1.5K. 
The operating temperature of 1.5K is held by decompressing ${}^{4}He$. 
Gain at the time of measurement was increased 1000 times. The noise was removed in RC circuit where 
cut off frequency of this circuit was 160Hz low-pass filter.

\begin{table}
\begin{center}
\begin{tabular}{l|l|l}
\hline
Source           & Black Body(1273K)         &IR-563\\
\hline
Source size      & A : 9mm,11mm, 13mm(trapezoid )&Masking for Al plate\\
		& B : 12mm, 20mm, 15mm		&	\\
		& C : 6mm, 10mm, 14mm		&	\\
		& D :6mm, 8mm, 10mm		&	\\
\hline
Base line length & A : 18cm - 28cm (2 - 6 steps), &  \\  
 & B- D : 18cm - 38cm (3 steps) &	  \\
\hline
Rotating angle   & A : 15 - 30degrees /steps   &		\\
		& B - D : 30 degrees /steps	&	\\	
\hline
Total baseline number    & A : 54, B : 20, C : 21, D : 18 &   \\
\hline
\end{tabular}
\end{center}
\caption{Parameters of Imaging experiment. Source A was measured in August 2003. and Source B - D were measured in February 2004. }
\label{2dP}
\end{table}

\section{RESULT}
\begin{figure}
   \begin{center}
   \begin{tabular}{c}
  \includegraphics[height=4cm]{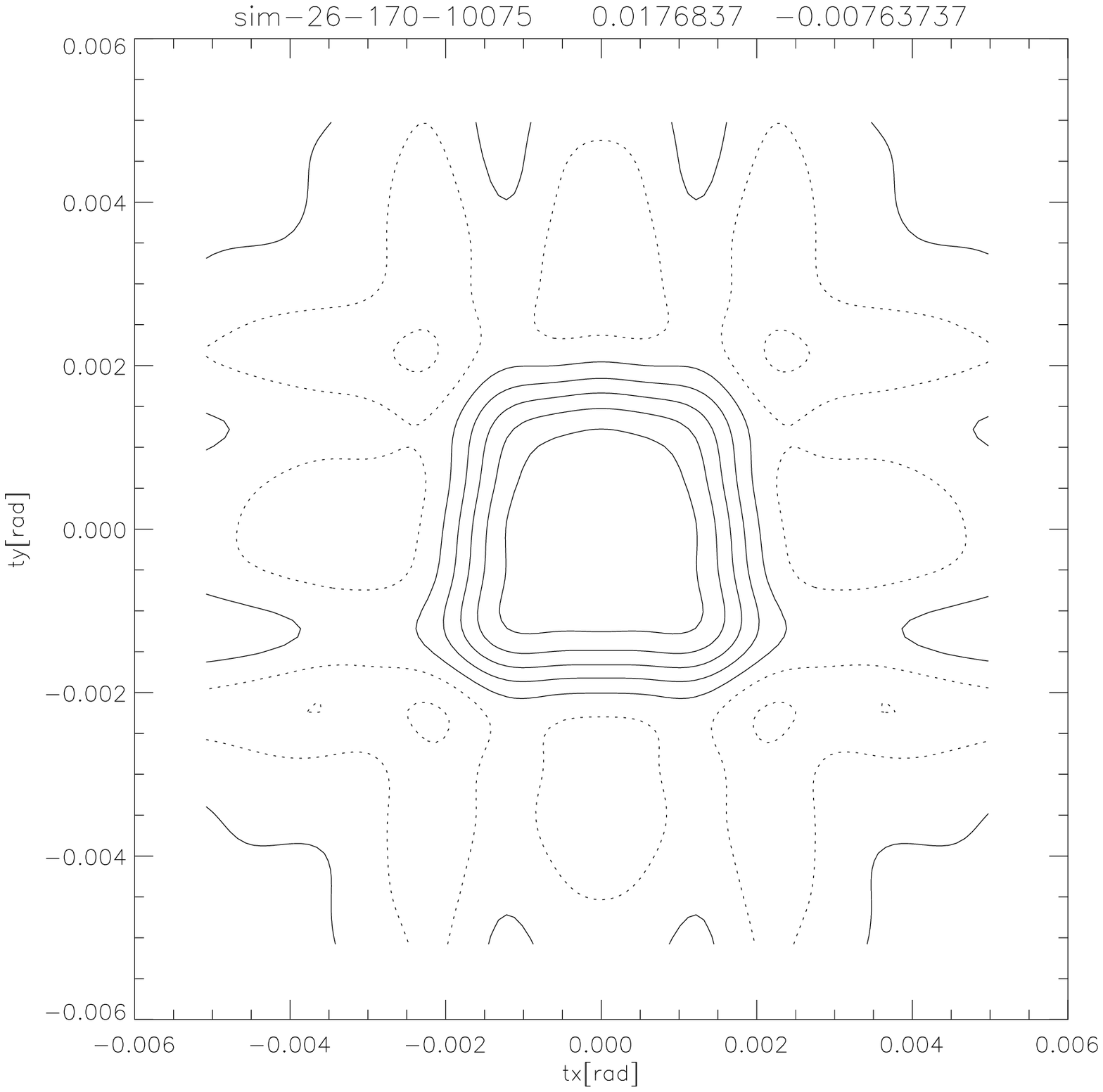}
   \includegraphics[height=4cm]{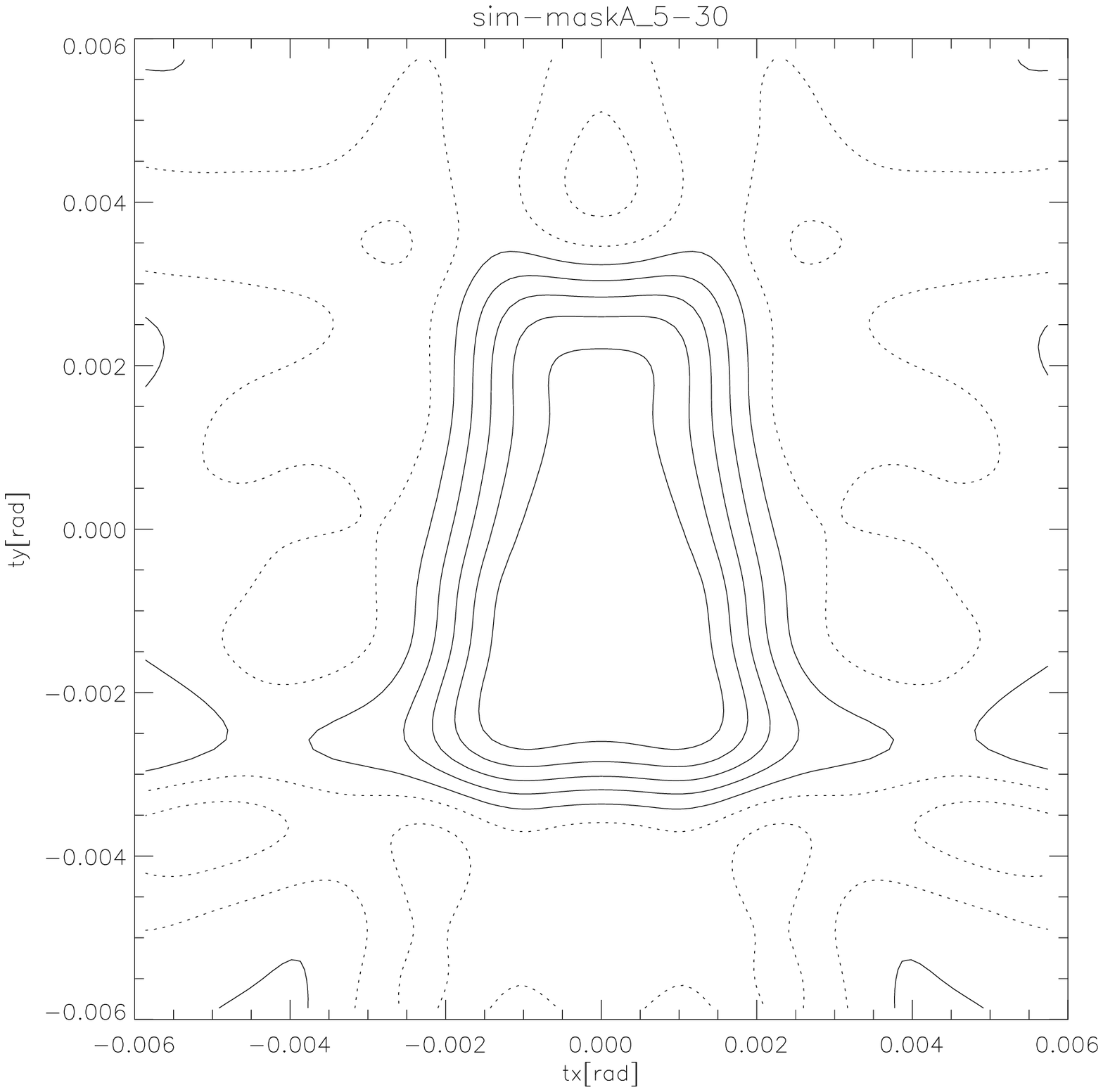}
   \includegraphics[height=4cm]{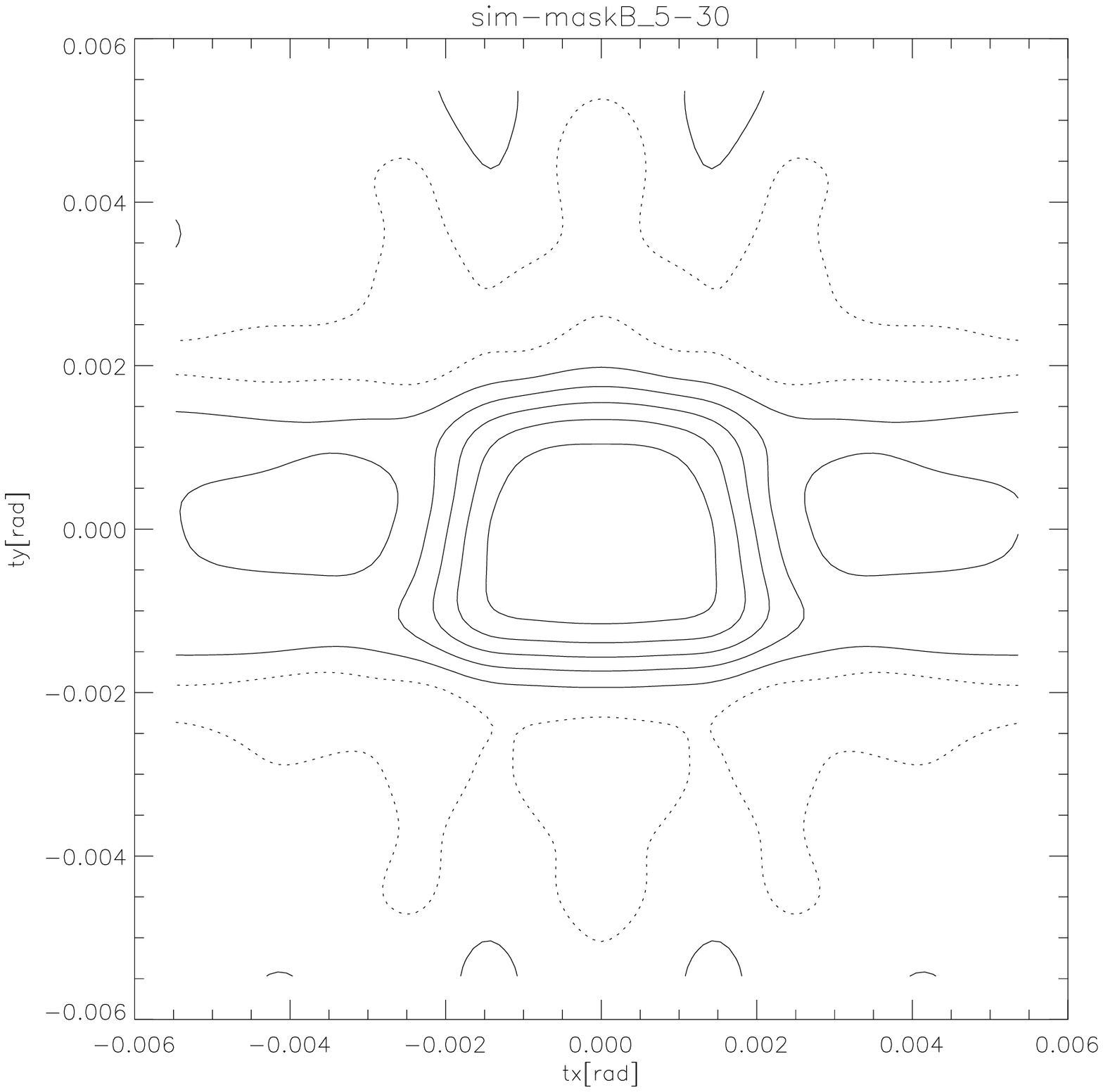}
   \includegraphics[height=4cm]{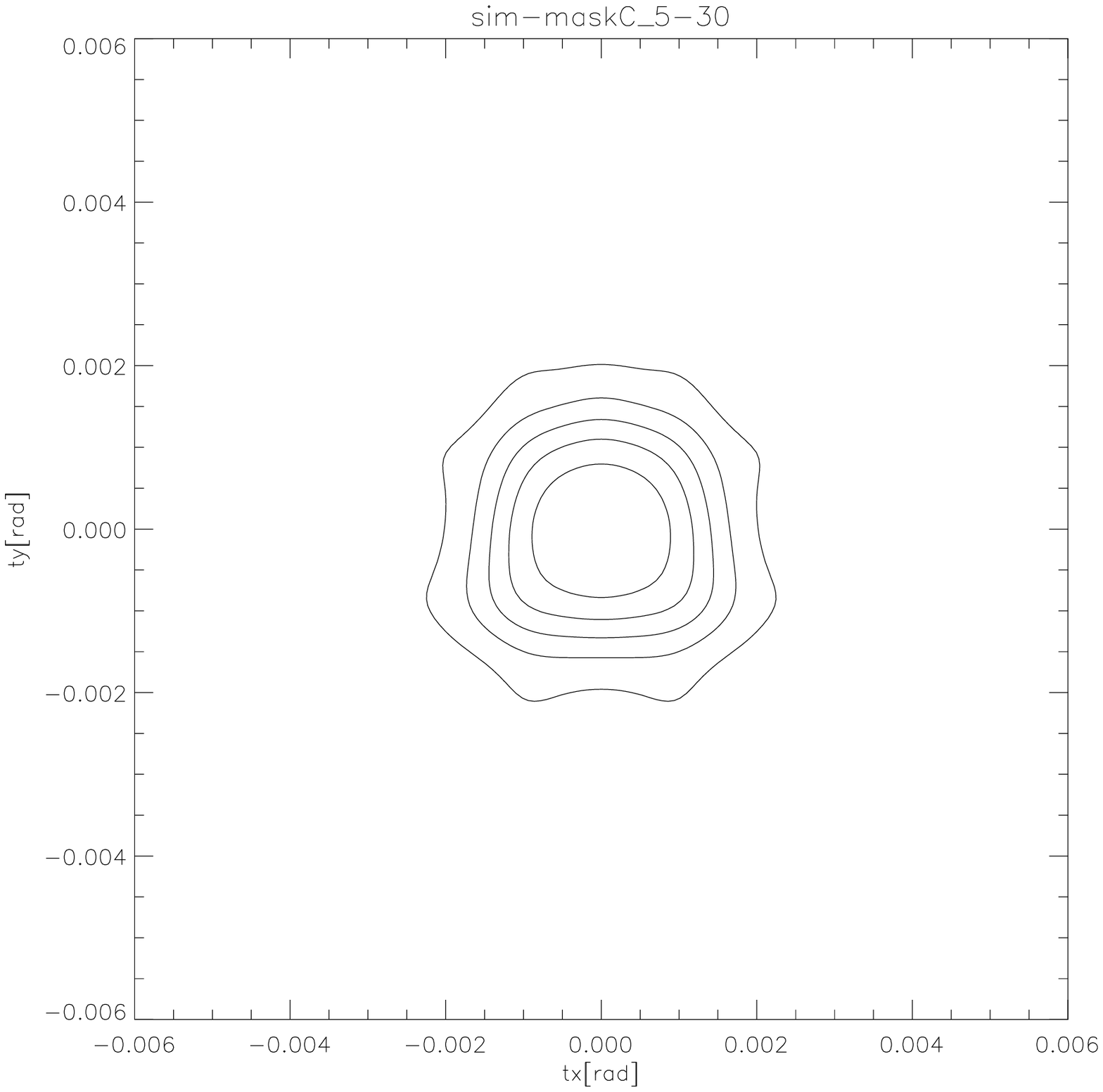}\\
  \includegraphics[height=4cm]{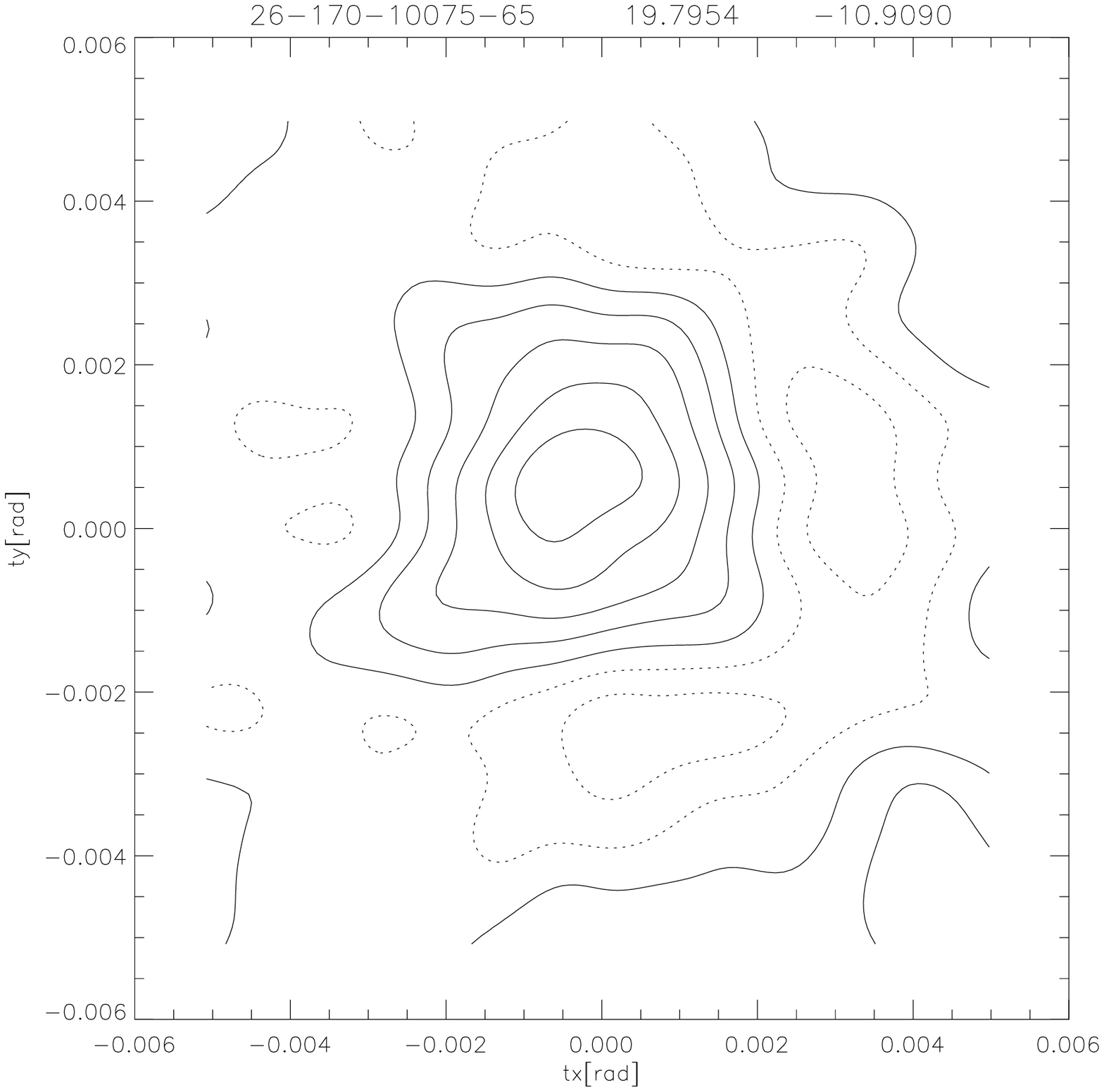}
     \includegraphics[height=4cm]{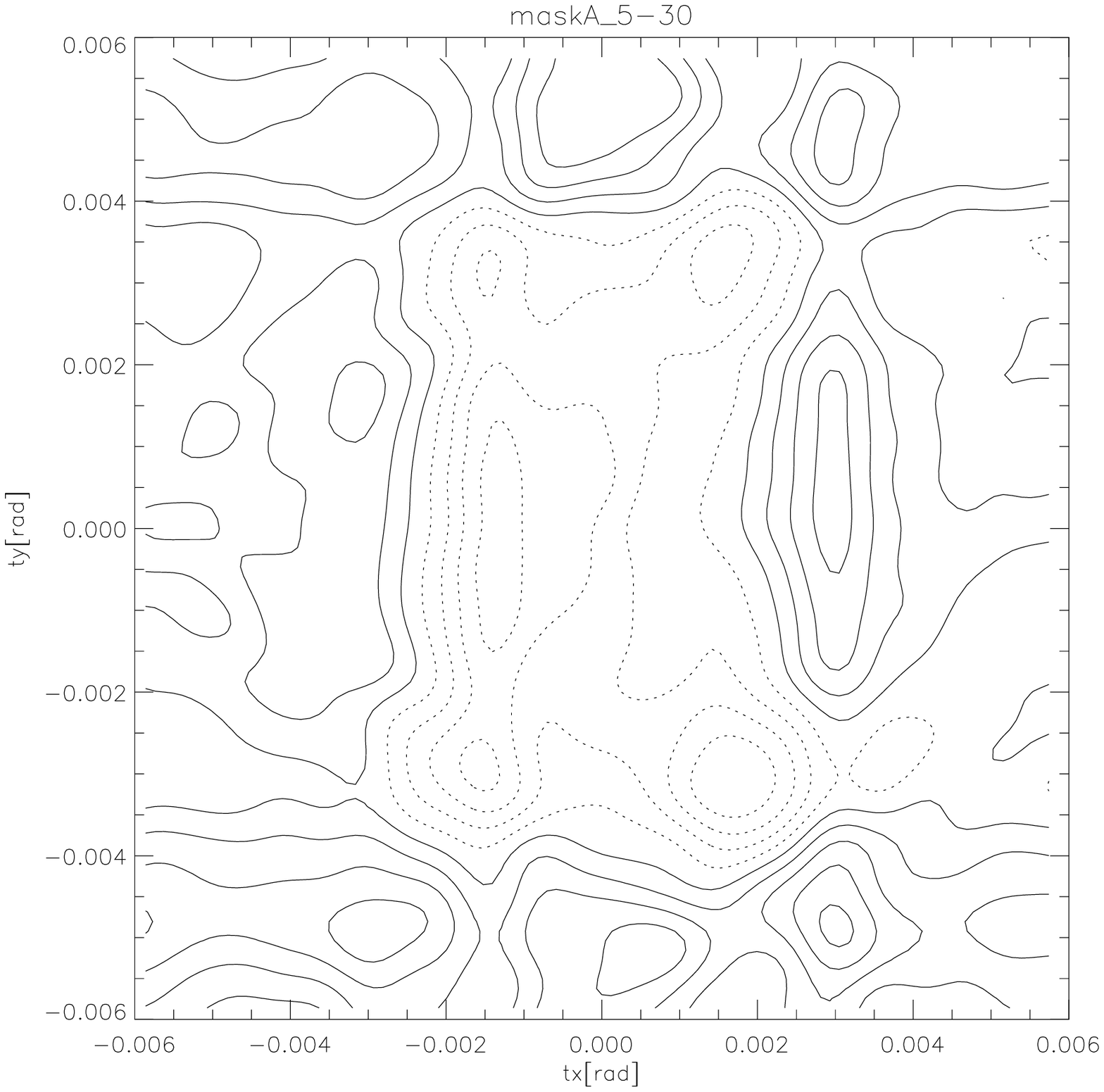}
   \includegraphics[height=4cm]{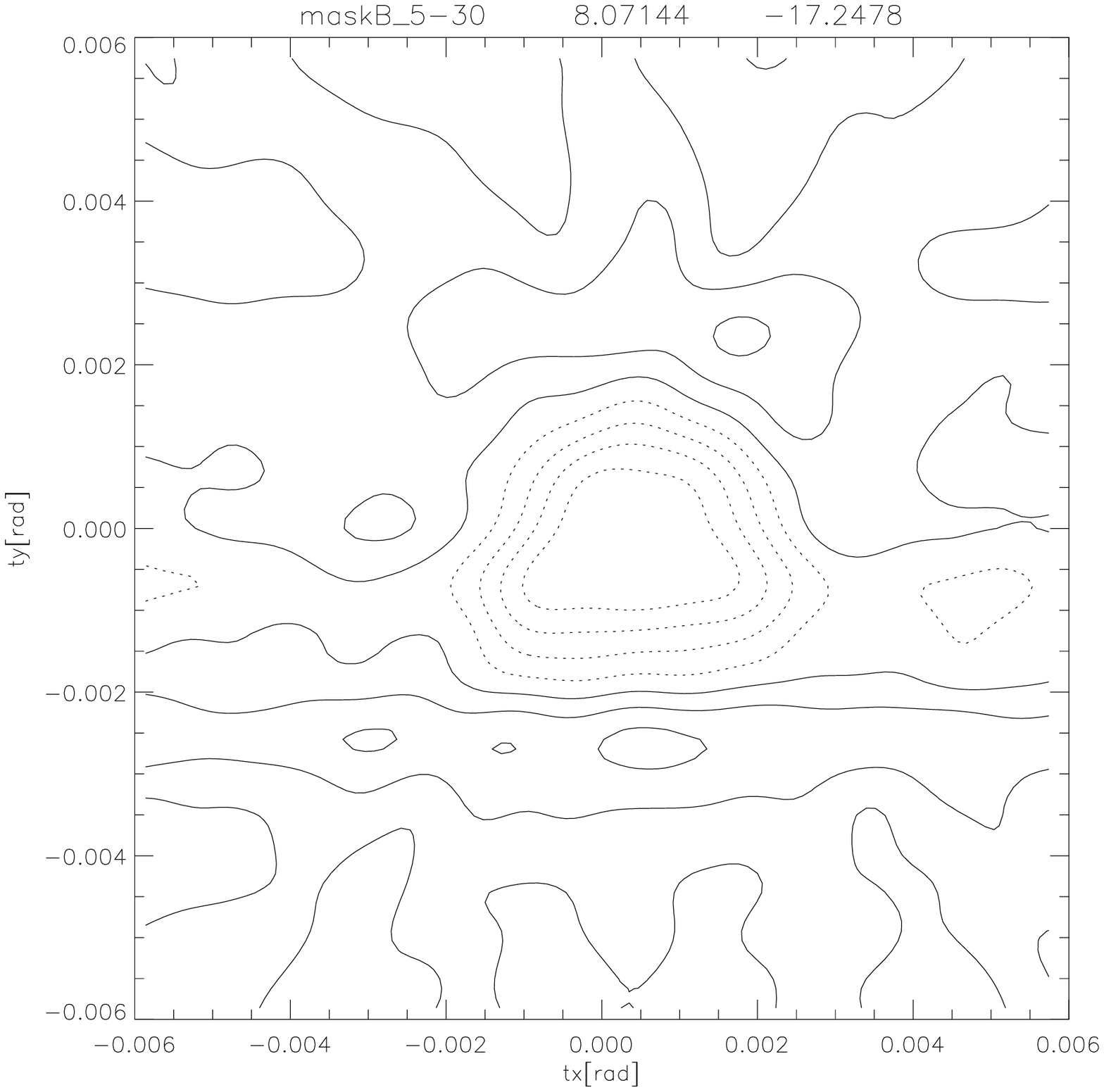}
   \includegraphics[height=4cm]{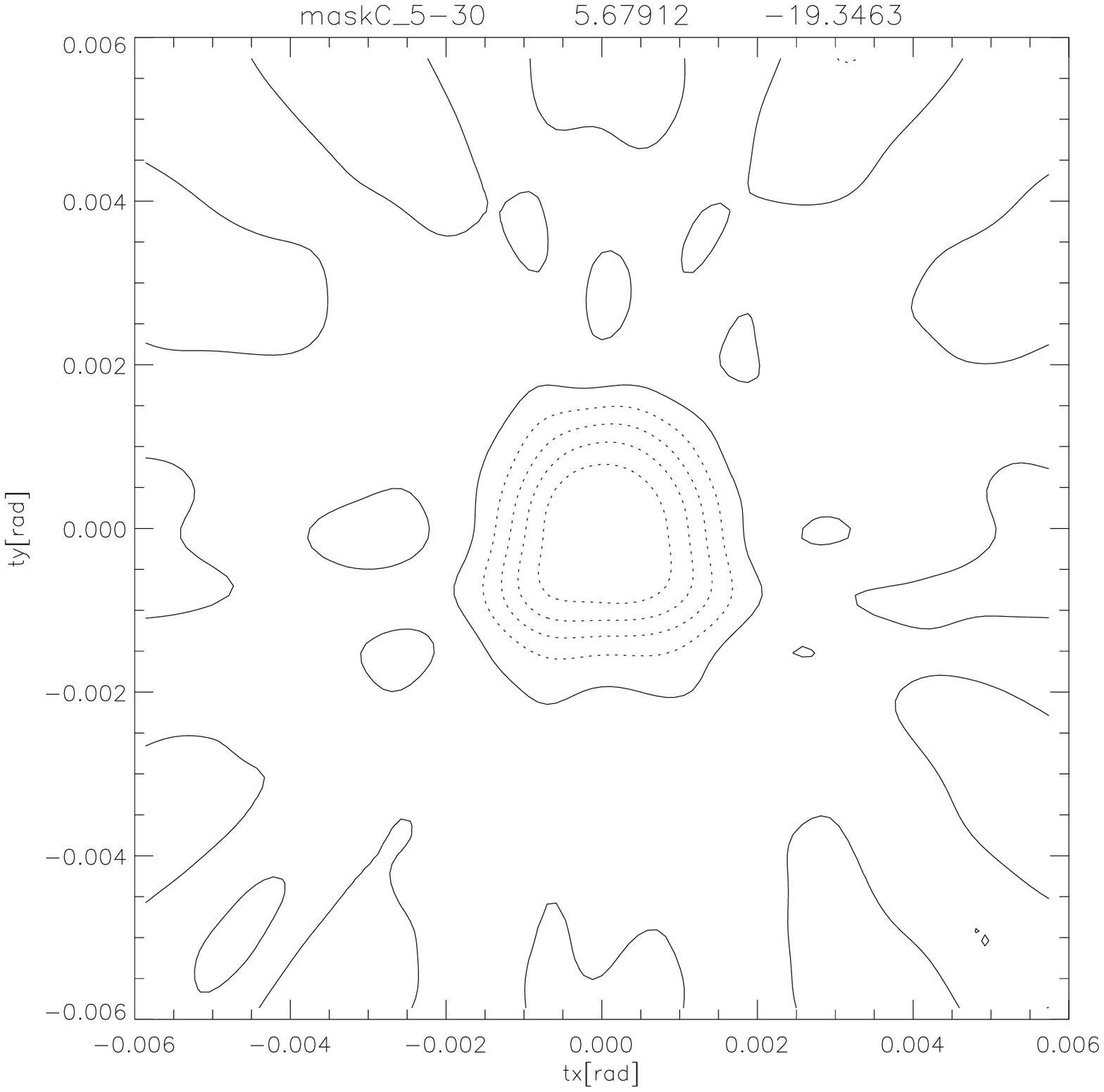}    
\end{tabular}
   \end{center}
   \caption{\label{fig:image1}Comparison of the simulated dirty images (top) and the observed dirty images (bottom) of masks A-D from left to right. All the obtained data for 5 - 30 cm$^{-1}$ are summed up into single images.   A vertical axis and a horizontal axis is $\theta _{x}, \theta_{y}$ in radian. }
\end{figure}

When the source intensity distribution does not depend on frequency, 
all images obtained for different frequencies are summed up into 
single image as discussed in the previous section. 
This is the large dynamic range merit of the MuFT.
To show this merit with concrete experimental results, 
observed dirty images of 4 mask patterns after all the data 
in 5 - 30 cm$^{-1}$ are summed up, are shown in Fig.\ref{fig:image1} 
bottom panels. The simulated dirty images for the same $u-v$ coverage 
and the same
frequency range are also shown for comparison in the top panels.
The simulated images are the expected images when the observations
were performed with noise free.

All of the observed images are well reproducing the simulated images.
These results have confirmed  
the capability of the simultaneous observations of 
source spectrum and images by the MuFT and the large dynamic range
of the MuFT.

Gain and NEP of the prototype MuFT were estimated from the above results. 
Gain of the system was estimated from the above results to be
$2.22\times 10^{-13}$[Volt$/$Jy] in the frequency ranges 
where the strong atmospheric 
absorption lines are absent. 
The system noise equivalent power was estimated to be $NEP_{system}\sim 
1.1\times 10^{-11}$[W/$\sqrt{Hz}$] based on the above experiments.
This rather high system NEP may be determined by the 
noise power of the atmospheric emission as explained below. 
NEP of this detector system $NEP_{det.}$is 
$1.0\times 10^{-14}/\eta$[W/$\sqrt{Hz}$] where $\eta$ is a quantum 
efficiency of the bolometer. This is much lower than the system NEP.
The NEP of 300 K atmospheric emission for 
this bolometer is about 2.3 $\times 10^{-13}[W/\sqrt{Hz}]$. The reasons  
why $NEP_{atomos.}$ takes such a high value are 
  a large FOV of 
the det. system and a high cut off frequency of the low path filter
mounted on the bolometer.
The measured FOV of det. system is about 
20$^{\circ}$.
The current detector system absorbs all radiation below 3THz. 
Due to the arrangement of the wire grids in option 1, 
only one eighth of the input source radiation is reached 
to the detector system at most and another 
radiation is transmitted and reflected out when the beams hit the WGs.
One the other hand, the noise atomospheric emission does not 
suffer this one eighth degradation since the everywhere around the 
bolometer is filled by 300K black body radiation in submm wavebands.  
The detection efficiency of detector system is about $\eta\sim 0.1 - 0.3$. 
Since the atmospheric noise emission was the dominant source of 
 the system noise, the expected system noise equivalent power 
 is  $NEP_{system}= 8/\sqrt{\eta} NEP_{atomos.}\sim =0.6\times 10^{-11}
 [W/\sqrt{Hz}] (\eta/0.1)^{-0.5}$. 
Although the obtained value is similar to the measured value, 
it  is still a factor 
of several smaller than the measured NEP of the system. 
The  may be able to attribute to the 
uncertainty in the FOV of the current detector and unknown quantum efficiency 
of the detector. A part of the discrepancy may 
also be due to still inaccurate adjustment of an optical system.
\section{SUMMARY \& DISCCUSION}

In this research, we have developed 
a new instrument by applying the aperture synthesis 
technique to the Fourier Transform Spectrometer (FTS) -- MuFT.
MuFT is a system which makes possible the imaging and
spectroscopy in a wide band by combining Wiener-Khinchine Formula 
which tells  that a spectrum can be measured by the auto-correlation, and van 
Citter Zernike Formula which makes imaging possible by the mutual 
correlation. 
Since the polarizers are used as the beam splitter and combiner 
in this equipment,  acquisition of polarization 
information is possible. Fundamentals of MuFT were considered, and
the concrete composition of this equipment is shown.

Imaging experiment of the extended source using the blackbody of about 1273 K was conducted.
Consequently, 
2D source images for various frequencies in the wide band from about 5cm$^{-1}$ to 30cm$^{-1}$
were successfully obtained.
Moreover, we have succeeded in reconstructing source image  by summing up all the images 
obtained at different frequencies and shown the large dynamic range merit experimentally.  

However, the following improvements are required for the optical system.
The adjustment accuracy of optical system is low.
For example, optical-axis adjustment with the light source 
and LiC carried out manually by observer's eyes. 
A certain device 
which improves the accuracy of this portion 
is desired for the usage in actual astronomical observation. 
The new optical system which improves these points 
needs to be 
designed and manufactured.
The base line length taken in this experiment 
was limited in the quite narrow range.
It is necessary to also perform measurement by the long baseline.
Moreover, we have not yet confirmed whether the same image can be 
observed by repeating the same experiment with the same mask and 
the same $u-v$ coverage.  
Actually, the phase of the obtained images for  the masks B-D are
shifted amount of $\pi$ from the expected images. 
The reason of this is still unresolved. 
Although all imaging experiments are conducted by trapezoid, 
imaging experiments with other types of the source, e.g. non-axis 
symmetric images, are also necessary.  

The system noise level was measured to be 
$NEP_{system}\simeq 1.1\times 10^{-11}[W/\sqrt{Hz}]$.
This is much higher than the intrinsic NEP of the bolometer of
$1.0\times 10^{-14}/$[W/$\sqrt{Hz}$].
The  NEP of 300 K atmospheric emission for this bolometer is about 
2.3 $\times 10^{-13}[W/\sqrt{Hz}]$.
By taking into account the efficiency of the optical system, that is less than
$1/8$, and the efficiency of the detector, that is about 0.1 to 0.3,  
discrepancy between the expected NEP of the 
system when the noise is limited by atmospheric emission and the measured NEP of 
the system,  is reduced to a factor of several. 
So the high noise level of the current system may be explained by the 
atmospheric emission and a large loss of the incident light due to 
very low efficiency of optical systems and detectors which still contain large
uncertainties. 

We have also shown that measurement of the Stokes parameters of 
$Q,U,V$ are possible for the point like quasimonochromatic wave source. 
However, for actually performing polarization observation, 
quantitive measurement of 
the Stokes parameters $\mathcal{Q, U}$, and $\mathcal{V}$
are required to be performed. 

In order to determine the accuracy of  imaging observations, 
the further supplementary examinations are required.
The FOV of the system was measured in this research.  
However, identification of the factors which limits FOV is now on. 

At B=18cm, 
$T_{int}\sim 80s$ and source size 10',the system is possible to detect a source of 
T$\sim$1300K and 10' at source size with $10\sigma$. 
Based on these results, we can estimate the observabilities by current system and improved 
system.
Using these results at S/N $\sim $ 10, source size 10', optical depth $\sim $ 1, and integration 
time 10 hours, the system is possible to detect a source  of T$\sim $ 166K. 
Sources which satisfy these conditions are  
restricted to quite bright objects, 
such as the sun and the moon.  While raising the observation experience value of 
this equipment using observation of these sources, we  should develop improved 
equipment. For example, if the aperture size is increased to 50cm from 5cm or   
$NEP_{system}$ is lowered down to  $10^{-13}$ level,  
plenty of objects could get into observable. 
When $NEP_{system} \sim 10^{-15}$ such as 0.3K bolometer system is achieved, 
the system is possible to detect SZE of COMA  galaxy cluster
in 5$\sigma $ in each wave-band with bandwidth of $\Delta \nu \sim 10$[GHz]
within an integration time of 
$t_{int}\sim 10^{5}$[s]. 

\section{FUTURE}

Finally, 
a future schedule is summarized. In a laboratory experiment, supplementary 
examination of two dimension imaging, estimate of accuracy, check the view 
for every frequency, identification of the noise origin, etc.  
Toward outdoor observations,  
LiC part must be corrected and optimized. 
Complete the development of 0.3K 
bolometer system. This improves the system NEP 
down to a $NEP_{system} \sim10^{-15} $ level. The source guiding system 
and tracking system must be constructed.

We are planning to perform the observations from  good mm and submm 
observation sites,
such as Nobeyama or Atacama. In Atacama, the atmospheric window 
from 200GHz to 300GHz is very attractive for MuFT. 
An interesting result is expected if spectrally resolved imaging observations 
of SZ in this wide range and 150GHz bands are performed.  
Since we can use the system without thinking of time sharing with 
many observers, 
original imaging and spectroscopic observation of nearby sources are
expected. 

We are planning to construct ground base 
mm and submm observation system based on MuFT 
consisted of  the aperture 
of 50cm telescopes, and maximum baseline length 
of a few  m.    
The possibility of taking extensive FOV can also be cultivated by the formation of 
many elements of bolometer, STJ, or TES. An optical design for that 
is also advanced. 

Applying this technique to space born mission is one of the best 
possibilities to extract the maximum ability of MuFT since 
there is no restriction on the band width from the atmospheric absorptions. 
The future
application of this technique to the observations from the space 
could open new interesting possibilities in FIR astronomy.

\acknowledgments     

The research was financially supported by the COE program
"Exploring New Science by Bridging Particle-Matter Hierarchy" from 
Tohoku University, the Sasagawa Scientific Research Grant from 
The Japan Science Society and by the Grant-in-Aid for Scientific
Research (116204010) of the Japan Society for the Promotion of
Science.


\bibliography{report1}   
\bibliographystyle{spiebib}   

\end{document}